\documentclass[pra,aps,superscriptaddress,floatfix,twocolumn]{revtex4-1}
\usepackage{amsmath,amssymb,multirow,epsfig,bm,pifont}
\usepackage[linkcolor=black,citecolor=black,urlcolor=blue,colorlinks=true,linktocpage=true]{hyperref}
\usepackage{graphicx}
\usepackage{epsfig}
\usepackage{bm}
\usepackage{tabularx}
\usepackage{makecell}

\newcommand{\beq}{\begin{equation}}
\newcommand{\eeq}{\end{equation}}
\newcommand{\bea}{\begin{eqnarray}}
\newcommand{\eea}{\end{eqnarray}}
\newcommand{\tr}{\mathrm{tr}}

\newcommand{\la}{\langle}
\newcommand{\ra}{\rangle}
\begin{document}

\title{Thermodynamics of rotating quantum matter in the virial expansion}
\author{C. E. Berger}
\affiliation{Department of Physics and Astronomy, University of North Carolina, Chapel Hill, NC, 27599, USA}
\author{K. J. Morrell}
\affiliation{Department of Physics and Astronomy, University of North Carolina, Chapel Hill, NC, 27599, USA}
\author{J. E. Drut}
\affiliation{Department of Physics and Astronomy, University of North Carolina, Chapel Hill, NC, 27599, USA}

\begin{abstract}
We characterize the high-temperature thermodynamics of rotating bosons and fermions in two- (2D) and 
three-dimensional (3D) isotropic harmonic trapping potentials.
We begin by calculating analytically the conventional virial coefficients $b_n$ for all $n$ in the noninteracting case,
as functions of the trapping and rotational frequencies. We also report on the virial coefficients for the angular 
momentum and associated moment of inertia.
Using the $b_n$ coefficients, we analyze the deconfined limit (in which the angular frequency matches the trapping frequency) and 
derive explicitly the limiting form of the partition function, showing from the thermodynamic standpoint how both the 2D and 3D cases 
become effectively homogeneous 2D systems.
To tackle the virial coefficients in the presence of weak interactions, we implement a coarse temporal lattice approximation 
and obtain virial coefficients up to third order. 
\end{abstract}

\date{\today}

\maketitle

%%%%%%%%%%%%%%%%%%%%%%%%%%%%
\section{Introduction}

The exploration of the phases of matter in regimes governed by quantum mechanics, i.e. quantum matter,
is now carried out with increasing accuracy and controllability in ultracold-atom experiments~\cite{Review1, Review2, ExpReviewLattices}.
The ability to tune the interaction strength via Feshbach resonances~\cite{ResonancesReview}, introduce imbalances such as mass and
polarization~\cite{Chevy2010}, vary the number of internal degrees of freedom, and control the temperature and external trapping potential, 
have led to a huge parameter space that experimentalists can realize and manipulate~\cite{RevExp}. 
These have in turn enabled a large body of work that continues to grow both qualitatively and quantitatively, toward elucidating the 
properties of quantum systems in extreme conditions as a function of internal as well as thermodynamic parameters.

Most notably, experiments already more than two decades ago achieved the first realizations of atomic Bose-Einstein 
condensates~\cite{BEC1995,PhysRevLett.75.3969} and about a decade later fermionic superfluids~\cite{PhysRevLett.92.040403,PhysRevLett.92.120403}, and since then experimentalists have continued to probe 
these systems in the various ways mentioned above and more. In particular, for both bosonic and fermionic systems, experimentalists early on realized rotating condensates and observed vortices and vortex lattices~\cite{Matthews, Madison, FermiVortices}, the latter widely regarded as the `smoking gun' for superfluidity.
From the condensed matter standpoint, the interest in rotating condensates is often associated with the realization of exotic strongly correlated states (such as those associated with the fractional quantum Hall effect; see e.g.~\cite{Cooper2008}). In those systems, the limit of large vortex number, i.e. large 
angular momentum, corresponds to the `deconfinement limit' in which the angular frequency matches the trapping frequency, and is of particular interest as it admits a simple description (in the case of weak interactions) in terms of Landau levels.

While there exists a considerable body of work on such rotating condensates (see e.g.~\cite{Cooper2008, RMPFetter} for reviews), i.e. work addressing the ground state and low-temperature phases, less is known about the specifics of the high-temperature behavior of these systems. In particular, little is known about the quantum-classical crossover and how strong correlations (which play a crucial role in determining the shape of the phase diagram~\cite{Stringari}) affect the normal phase of rotating strongly coupled matter. 

In this work we provide another piece of the puzzle by analyzing the high-temperature thermodynamics of rotating Bose and 
Fermi gases in 2D and 3D. To that end, we use the virial expansion and implement a coarse temporal lattice approximation recently put forward in Refs.~\cite{ShillDrut,HouEtAl,MorrellEtAl}. 
The approximation allows us to bypass the requirement of solving the $n$-body problem 
to access the $n$-th order virial coefficient, which will be essential to address the effects of interactions. For the sake of simplicity, we will furthermore focus on systems with two 
particle species with a contact interaction across species (i.e. no intra-species interaction). Along the way, we present in detail several results for noninteracting systems which, while easy to obtain and should be textbook material, do not appear in the literature to the best of our knowledge. Previous work addressing the high-temperature thermodynamics of rotating 
quantum gases, e.g. in interacting~\cite{Mulkerin2012_1, Mulkerin2012_2} as well as noninteracting~\cite{Li2016, LiGu2016} regimes, present different analyses which are complementary to the present work.

%%%%%%%%%%%%%%%%%%%%%%%%%%%%
\section{Hamiltonian and formalism}

As our focus is on systems with short-range interactions, such as dilute atomic gases or dilute neutron matter, 
the Hamiltonian reads
\beq
\hat H = \hat H_0 +  \hat V_\text{int},
\eeq
where
\beq
\hat H_0 = \hat T + \hat V_\text{ext} - \omega_z \hat L_z,
\eeq
and
\bea
\label{Eq:T}
\hat T \!=\! \sum_{s = 1,2} {\int{d^d x\,\hat{\psi}^{\dagger}_{s}({\bf x})\left(-\frac{\hbar^2\nabla^2}{2m}\right)\hat{\psi}_{s}({\bf x})}},
\eea
is the kinetic energy,
\bea
\label{Eq:Vext}
\hat V_\text{ext} \!=\! \frac{1}{2} m \omega_\text{tr}^2 \int{d^d x}\, {\bf x}^2 \,( \hat{n}_{1}({\bf x}) + \hat{n}_{2}({\bf x})) ,
\eea
is the spherically symmetric external trapping potential,
\bea
\label{Eq:Vint}
\hat V_\text{int} \!=\! - g_{d}\! \int{d^d x\,\hat{n}_{1}({\bf x})\hat{n}_{2}({\bf x})},
\eea
is the interaction, and
\bea
\label{Eq:Lz}
\hat L_z \!=\! -i\sum_{s = 1,2}{\int{d^d x\,\hat{\psi}^{\dagger}_{s}({\bf x})\left(x \partial_y - y \partial_x \right)\hat{\psi}_{s}({\bf x})}},
\eea
is the angular momentum operator in the $z$ direction. In polar or spherical coordinates, the differential operator in the above
second-quantized form becomes simply $-i \partial/\partial \phi$ where $\phi$ is the azimuthal angle.
In the above equations, the field operators $\hat{\psi}_{s}, \hat{\psi}^{\dagger}_{s}$ correspond to particles of species $s=1,2$, 
and $\hat{n}_{s}({\bf x})$ are the coordinate-space densities. 
In the remainder of this work, we will take $\hbar = k_\text{B} = m = 1$.

%%%%%%%%%%%%%%%%%%%%%%%%%%%%
\subsection{Thermodynamics and the virial expansion}

The equilibrium thermodynamics of our quantum many-body system is captured by the grand-canonical partition function,
namely
\beq
\mathcal Z = \tr \left[ e^{-\beta (\hat H -\mu \hat N)}\right] = e^{-\beta \Omega},
\eeq
where $\beta$ is the inverse temperature, $\Omega$ is the grand thermodynamic potential, $\hat N$ is the total particle number operator, 
and $\mu$ is the chemical potential for both species.

At this point, it is useful to review the parameters that control our system, including the thermodynamic ones; 
they are: $\beta$, $\mu$, $\omega_\text{tr}$, $\omega_z$, and $g_{d}$. We may then form dimensionless parameters, which
we may choose to be $\beta \mu$, $\beta \omega_\text{tr}$, $\beta \omega_z$, and $\lambda$, where the latter will typically involve a
scattering length and will depend on whether we are examining the 2D or 3D problems (see below).

As the calculation of $\mathcal Z$ is a formidable problem in the presence of interactions, we resort to approximations and numerical
evaluations in order to access the thermodynamics. To that end, in this work we will explore the virial expansion (see Ref.~\cite{VirialReview} for a review), which is an expansion around the dilute limit
$z\to 0$, where $z=e^{\beta \mu}$ is the fugacity, i.e. it is a low-fugacity expansion. The coefficients accompanying the powers of $z$ in 
the expansion $\Omega$ are the virial coefficients $b_n$:
\beq
-\beta \Omega = \ln {\mathcal Z} = Q_1 \sum_{n=1}^{\infty} b_n z^n,
\eeq
where $Q_1$ is the one-body partition function. Using the fact that $\mathcal Z$ is itself a 
sum over canonical partition functions $Q_N$ of all possible particle numbers $N$, namely
\beq
\mathcal Z = \sum_{N=0}^{\infty} z^N Q_N,
\eeq 
we obtain expressions for the virial coefficients
\bea
b_1&=& 1,\\
b_2 &=& \frac{Q_2}{Q_1} - \frac{Q_1}{2!},\\
b_3 &=& \frac{Q_3}{Q_1} - b_2 Q_1  - \frac{Q_1^2}{3!},
\eea
and so on. In this work we will not pursue the virial expansion beyond $b_3$. 
The $Q_N$ can themselves be written in terms of the partition functions $Q_{a,b}$ for $a$ particles of type 1 and $b$ particles of type 2:
\bea
Q_1 &=& 2 Q_{1,0},\\
Q_2 &=&  2 Q_{2,0} + Q_{1,1},\\
Q_3 &=&  2 Q_{3,0} + 2 Q_{2,1},
\eea
and so on for higher orders. In the absence of intra-species interactions, only the $Q_{1,1}$ and $Q_{2,1}$ are affected, such that
the change in $b_2$ and $b_3$ due to interactions is entirely given by
\bea
\label{Eq:Db2Db3}
\Delta b_2 &=& \frac{\Delta Q_{1,1}}{Q_1}, \\
\Delta b_3 &=& \frac{2 \Delta Q_{2,1}}{Q_1} - \Delta b_2 Q_1.
\eea

We will use these expressions to access the high-temperature thermodynamics of bosons and fermions. To calculate $\Delta Q_{1,1}$
and $\Delta Q_{2,1}$, we will implement a coarse temporal lattice approximation, as described in the next section. Once we obtain the virial 
coefficients, we will rebuild the grand-canonical potential $\Omega$ to access the thermodynamics of the system as a function of
the various parameters. In order to connect to the {\it physical} parameters of the systems at hand, one may use the value of $\Delta b_2$
as a renormalization condition by relying on the exact answer, which is known at $\omega_z = 0$; namely,
\bea
\Delta b_2^\text{(2D)} &=& \frac{e^{-\beta \omega_\text{tr}}}{2} \sum_{n=0}^{\infty}\left[ e^{-\beta \omega_\text{tr} 2 \nu_n (\lambda)} - e^{-\beta \omega_\text{tr} 2 n} \right ], \\
\Delta b_2^\text{(3D)} &=& \frac{e^{-\beta \omega_\text{tr} 3 / 2}}{2} \sum_{n=0}^{\infty}\left[ e^{-\beta \omega_\text{tr} 2 \nu_n (\lambda)} - e^{-\beta \omega_\text{tr} 2 n} \right ].
\eea
[see Ref.~\cite{Exact2DHu} for the 2D case and~\cite{BuschEtAl} for the 3D case], where
$\omega_\text{tr} (2 \nu_n (\lambda) + d/2)$ is the energy of the $d$-dimensional two-body problem in the center-of-mass frame.
Using these expressions, one may fix the value of the dimensionless coupling for each system, for a given $\beta \omega_\text{tr}$.
The use of $\Delta b_2$ as a physical quantity to renormalize the coupling constant was advocated in Refs.~\cite{ShillDrut,HouEtAl,MorrellEtAl}.

%%%%%%%%%%%%%%%%%%%%%%%%%%%
\subsection{Single-particle basis and single-particle partition function in 2D and 3D}

In evaluating the results of the coarse temporal lattice approximation presented below, we will use the eigenstates of 
$\hat H_0$ in 2D and 3D, in polar and spherical coordinates, respectively. We therefore present them in detail here for future
reference, along with the corresponding single-particle partition function.

\subsubsection{Two spatial dimensions} 

In 2D, the single-particle eigenstates of $\hat H_0$ in 2D are given by
\beq
\label{Eq:wf2D}
\langle {\bf x} | {\bf k} \rangle = \frac{1}{\sqrt{2\pi}} R_{k m}(\rho) e^{-i m \phi},
\eeq
where
\beq
R_{k m}(\rho) =  N_{km}^{(\text{2D})}{\omega^{1/2}_\text{tr}}\; e^{-\rho^{2}/2} \rho^{|m|}L_{k}^{|m|} (\rho^{2}),
\eeq
where $\rho = {\omega^{1/2}_\text{tr}}\; r$, and
\beq
N_{km}^{(\text{2D})} =  \sqrt{2}\sqrt{\frac{k!}{(k + |m|)!}},
\eeq
with $L_{k}^{|m|}$ the associated Laguerre functions.
We have used polar coordinates $r,\phi$, and a collective quantum number ${\bf k} = (k,m)$,
with $k = 0,1,\dots$ and $m$ can take any integer value. The corresponding energy is
\beq
E_{km} = \omega_\text{tr} (2k + |m| + 1) + \omega_z m .
\eeq

With this spectrum, it is a simple matter to calculate $Q_{1}$, which by definition is
\beq
Q_1 = \sum_{{\bf k}} e^{-\beta E_{\bf k}}.
\eeq

Thus, in 2D, 
\beq
\label{Eq:Q1_2D}
Q_1 =  2\sum_{k,m} e^{- \beta E_{k m}} = \frac{2\, e^{-\beta \omega_\text{tr}}} {(1 - e^{-\beta \omega_+}) (1 - e^{-\beta \omega_-})},
\eeq
where $\omega_\pm = \omega_\text{tr} \pm  \omega_z$ and the overall factor of 2 reflects the fact that
we have two particle species.

\subsubsection{Three spatial dimensions}

In 3D, the single-particle eigenstates of $\hat H_0$ in 3D are
\beq
\label{Eq:wf3D}
\langle {\bf x} | {\bf k} \rangle = R_{kl}(\rho) P_l^{m} (\cos \theta) e^{-i m \phi},
\eeq
where $P_l^{m} (x)$ are the associated Legendre functions and
\beq
R_{k l}(\rho) = N_{k l}^{(\text{3D})} \omega^{3/4}_\text{tr} e^{-\rho^{2}/2} \rho^l L_{k}^{l + 1/2} (\rho^{2}) ,
\eeq
where
\beq
N_{kl}^{(\text{3D})} =  \sqrt{\frac{1}{\sqrt{4 \pi}} \frac{2^{k + 2l + 3}\, k!}{(2k + 2l + 1)!!}}.
\eeq
Here, we have used spherical coordinates $r,\theta,\phi$, where $\theta$ is the polar angle, and $\phi$ the azimuthal angle. The 
collective quantum number ${\bf k} = (k,l,m)$ is such that $k \geq 0$, $l \geq 0 $, and $-l \leq m \leq l$.
The corresponding energy is
\beq
E_{k l m} = \omega_\text{tr} (2k + l + 3/2) + \omega_z m.
\eeq

Here, the corresponding single-particle partition function is given by
\bea
\label{Eq:Q1_3D}
Q_1 &=& 
\frac{2e^{-\beta \omega_\text{tr} 3/2} }{(1 - e^{-\beta \omega_\text{tr}}) (1 - e^{-\beta \omega_{+}}) (1 - e^{-\beta \omega_{-}})}.
\eea
%

%%%%%%%%%%%%%%%%%%%%%%%%%%%%
\subsection{Coarse temporal lattice approximation}

To calculate the interaction-induced change in the canonical partition functions $\Delta Q_{1,1}$ and $\Delta Q_{2,1}$, we 
propose an approximation which consists in keeping only the leading term in the Magnus expansion:
\beq
e^{-\beta (\hat H_0 + \hat V_\text{int})} = e^{-\beta \hat H_0}e^{-\beta \hat V_\text{int}}\times e^{-\frac{\beta^2}{2} [\hat H_0, \hat V_\text{int}]}\times \dots,
\eeq
where the higher orders involve exponentials of nested commutators of $\hat H_0$ with $\hat V_\text{int}$. Thus, the LO in this
expansion consists in setting $[\hat H_0 , \hat V_\text{int}] = 0$, which becomes exact in the limit where either $\hat H_0$
or $\hat V_\text{int}$ can be ignored (i.e. respectively the strong- and weak-coupling limits).
Previous explorations of this approximation, by us and others~\cite{ShillDrut, HouEtAl, MorrellEtAl, HouDrut}, indicate that LO-level results
(the so-called semiclassical approximation) for trapped systems are
not only qualitatively but also quantitatively correct at weak coupling.

%%%%%%%%%%%%%%%
\subsubsection{Two-body contribution $\Delta Q_{1,1}$.}

To calculate $\Delta b_2$ we will need the above result for $Q_1$ but also $\Delta Q_{1,1}$.
At leading order in our coarse temporal lattice approximation,
\bea
Q_{1,1} &=& \tr_{1,1}\left[e^{-\beta\hat{H}_0}e^{-\beta\hat{V}_\text{int}}  \right]  \nonumber \\
&=& \sum_{{\bf k}_{1},{\bf k}_{2},{\bf x}_{1},{\bf x}_{2}}\la {\bf k}_{1} {\bf k}_{2} | e^{-\beta \hat{H}_0} | {\bf x}_{1}{\bf x}_{2} \ra \la {\bf x}_{1}{\bf x}_{2} | e^{-\beta\hat{V}_\text{int}} |{\bf k}_{1} {\bf k}_{2} \ra \nonumber \\
&=& \sum_{{\bf k}_{1},{\bf k}_{2},{\bf x}_{1},{\bf x}_{2}} e^{-\beta(E_{{\bf k}_{1}} + E_{{\bf k}_{2}})} M_{{\bf x}_{1},{\bf x}_{2}} |\la {\bf k}_{1} {\bf k}_{2} | {\bf x}_{1}{\bf x}_{2} \ra|^{2},
\eea
where we have inserted complete sets of states in coordinate space $\{| {\bf x}_1 {\bf x}_2 \rangle\}$ and in the basis $\{ |{\bf k}_1 {\bf k}_2 \rangle \}$ of eigenstates of 
$\hat H_0$, whose single-particle eigenstates $| {\bf k} \rangle$ have eigenvalues $E_{\bf k}$.
We have also made use of the fact that $\hat V_\text{int}$ is diagonal in coordinate space, such that
\beq
M_{{\bf x}_{1},{\bf x}_{2}} = 1 + C \ell^{-d}\delta_{{\bf x}_1,{\bf x}_2},
\eeq
where $C = \ell^d \left(e^{\beta g_{d}} - 1\right)$ and we have introduced a spatial lattice spacing $\ell$ as a regulator.

Thus,
\beq
\Delta Q_{1,1} = C \sum_{{\bf k}_{1},{\bf k}_{2},{\bf x}} \ell^d e^{-\beta(E_{{\bf k}_{1}} + E_{{\bf k}_{2}})} |\la {\bf k}_{1} {\bf k}_{2} | {\bf x}\, {\bf x} \ra|^{2}.
\eeq

\begin{figure}[t]
\includegraphics[width=\columnwidth]{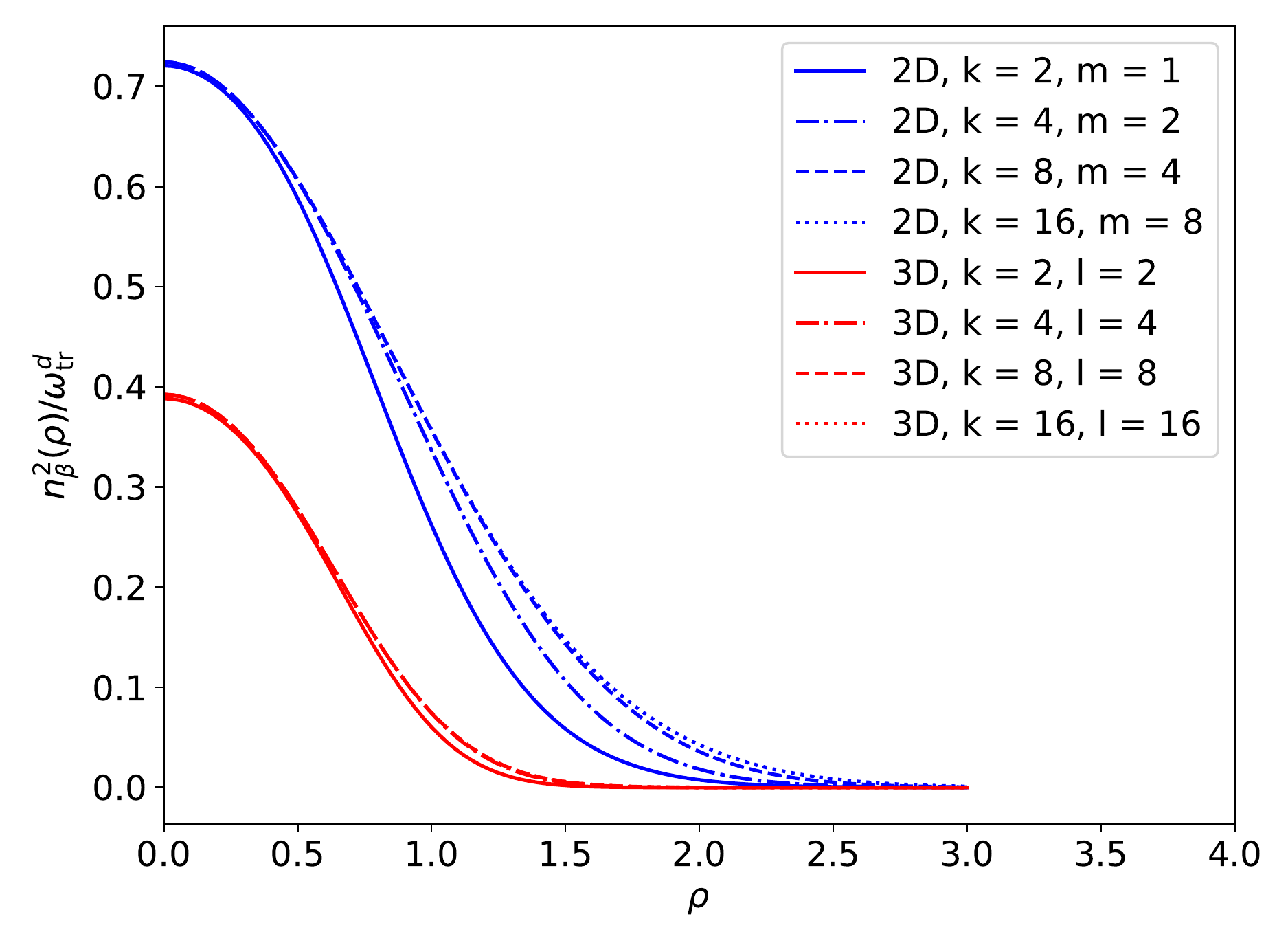} 
\caption{\label{Fig:CutoffEffects} $n_\beta^2({\bf x})/\omega_\text{tr}^d$ as a function of the radial coordinate $\rho$, for several
cutoff values of the quantum numbers $k,m$ in 2D (blue) and $k,l$ in 3D (red); in the latter case, the quantum number $m$ is summed over its full range $[-l,l]$. In this plot, $\omega_z/\omega_\text{tr}= 1/2$.}
\end{figure}

The computationally demanding part of this calculation is the overlap function $|\la {\bf k}_{1} {\bf k}_{2} | {\bf x} \, {\bf x} \ra|^{2}$. 
In this particular case, i.e. for $\Delta Q_{1,1} $, the overlap function can be factorized as $|\la {\bf k}_{1} | {\bf x} \ra |^2 | \la {\bf k}_{2} | {\bf x} \ra |^{2}$. Upon summing over ${\bf k}_1, {\bf k}_2$, we obtain a simpler expression
\beq
\Delta Q_{1,1} = C \sum_{{\bf x}} \ell^d n_\beta^2({\bf x})
\eeq
where
\beq
n_\beta({\bf x}) = \sum_{{\bf k}} e^{-\beta E_{\bf k}} |\la {\bf k} | {\bf x} \ra|^{2}.
\eeq
The exponential decay with the energy will enable us to cut off the sum over $\bf k$ without significantly losing precision.
We show a representative example of such cutoff effects in Fig.~\ref{Fig:CutoffEffects}.

Notice that $n_\beta({\bf x})$ has units of $\omega_\text{tr}^{d/2}$ [which corresponds to $(\text{length})^{-d}$] 
and it is a function of the dimensionless ratio $\rho =  {\omega^{1/2}_\text{tr}}\; r$ (see below for 2D and 3D examples), where $r = |{\bf x}|$.
Upon taking the continuum limit, 
\beq
\Delta Q_{1,1} \to \frac{C}{\lambda_T^d} \int d^{d} {\bf \bar x}\; {(2\pi \beta \omega_\text{tr})}^{d/2} \, \frac{n_\beta^2({\bf \bar x})}{\omega_\text{tr}^{d}},
\eeq
where ${\bf \bar x}  =  {\omega^{1/2}_\text{tr}}\; {\bf x}$ is dimensionless, and

Thus, in 2D,
\beq
n_\beta({\bf x}) = \omega_\text{tr} \frac{e^{-\rho^{2}}}{2\pi} 
\sum_{k,m} e^{-\beta E_{km}} f^{\text{2D}}_{km}(\rho^2),
\eeq
whose units come from the prefactor $\omega_\text{tr}$ and, as expected from symmetry considerations, is only a function of the radial coordinate (concentric with the trapping potential). Here,
\beq
f^{\text{2D}}_{km}(\rho^2) \equiv \frac{2\, k!}{(k + |m|)!} \rho^{2|m|}\left( L_{k}^{|m|} (\rho^{2})\right)^2,
\eeq

Similarly, in 3D,
\beq
n_\beta({\bf x}) = \omega_\text{tr}^{3/2} \frac{e^{-\rho^{2}}}{\sqrt{4\pi}} 
\sum_{k,l,m} e^{-\beta E_{klm}} f^{\text{3D}}_{kl}(\rho^2) (P^m_l (\cos \theta))^2
\eeq
where
\beq
f^{\text{3D}}_{kl}(\rho^2) \equiv \frac{2^{k + 2l + 3}\, k!}{(2k + 2l + 1)!!} \rho^{2l} \left (L_{k}^{l + 1/2} (\rho^{2}) \right)^2
\eeq

Using the above results, together with Eq.~(\ref{Eq:Db2Db3}) for $\Delta b_2$, we solve for the dimensionless quantity $B/\lambda_T^d$ in terms
of $\Delta b_2$:
\beq
\frac{C}{\lambda_T^d} = \Delta b_2 \frac{Q_1}{(2\pi \beta \omega_\text{tr})^{d/2}} \left({\int d^{d} {\bf \bar x}\;  \, \frac{n_\beta^2({\bf \bar x})}{\omega_\text{tr}^{d}}}\right)^{-1}.
\eeq

%%%%%%%%%%%%%%
\subsubsection{Three-body sector: $\Delta Q_{2,1}$ for fermions}

Following the same steps outlined above, it is straightforward to show that
\bea
\Delta Q_{2,1} &=& \frac{C}{2}\sum_{{\bf k}_{1}{\bf k}_{2}{\bf k}_{3}}e^{-\beta(E_{{\bf k}_{1}} + E_{{\bf k}_{2}} + E_{{\bf k}_{3}})}  \nonumber \\
&& \ \ \times \sum_{{\bf x}_{1}{\bf x}_{2}}| \la {\bf x}_{1}{\bf x}_{2}{\bf x}_{1} | {\bf k}_{1}{\bf k}_{2}{\bf k}_{3}\ra |^{2}.
\eea

The overlap can be simplified slightly by factoring across distinguishable species: 
\bea
\la {\bf x}_{1} {\bf x}_{2} {\bf x}_{1} | {\bf k}_{1} {\bf k}_{2} {\bf k}_{3} \ra &=& \la {\bf x}_{1} {\bf x}_{2}  | {\bf k}_{1} {\bf k}_{2} \ra \la {\bf x}_{1}  | {\bf k}_{3} \ra,
\eea
where the matrix element $\la {\bf x}_{1} {\bf x}_{2}  | {\bf k}_{1} {\bf k}_{2} \ra$ is a Slater determinant of single-particle states:
\beq
\la {\bf x}_{1} {\bf x}_{2}  | {\bf k}_{1} {\bf k}_{2} \ra = \la {\bf x}_{1} | {\bf k}_{1} \ra \la {\bf x}_{2} | {\bf k}_{2} \ra - \la {\bf x}_{2} | {\bf k}_{1} \ra \la {\bf x}_{1} | {\bf k}_{2} \ra.
\eeq

As in the case of $\Delta Q_{1,1}$, we will sum over the energy eigenstates first, and then perform the spatial sum.
To that end, it is useful to define
\beq
n^{F}_\beta({\bf x}_1,{\bf x}_2) = n_\beta({\bf x}_1) \sum_{{\bf k}_{1}{\bf k}_{2}} e^{-\beta (E_{{\bf k}_{1}} + E_{{\bf k}_{2}})} |\la {\bf x}_{1} {\bf x}_{2}  | {\bf k}_{1} {\bf k}_{2} \ra |^{2}
,
\eeq
such that,
\bea
\Delta Q_{2,1} &=& \frac{C}{2}\sum_{{\bf x}_{1}{\bf x}_{2}}n^{F}_\beta({\bf x}_1,{\bf x}_2).
\eea
As in the case of $n_\beta({\bf x})$, the exponential decay with the energy allows us to cutoff the double sum in $n^{F}_\beta({\bf x}_1,{\bf x}_2)$ without
significantly affecting the precision of the whole calculation.

%%%%%%%%%%%%%%
\subsubsection{Three-body sector: $\Delta Q_{2,1}$ for bosons}

The bosonic case differs from the fermionic case in that we must use a permanent rather than a Slater determinant. Thus, 
\beq
n^{B}_\beta({\bf x}_1,{\bf x}_2) =  n_\beta({\bf x}_1) \sum_{{\bf k}_{1}{\bf k}_{2}} e^{-\beta (E_{{\bf k}_{1}} + E_{{\bf k}_{2}})} |\la {\bf x}_{1} {\bf x}_{2}  | {\bf k}_{1} {\bf k}_{2} \ra |^{2},
\eeq
where the two-body overlap is now symmetric in its arguments, as befits bosons:
\beq
\la {\bf x}_{1} {\bf x}_{2}  | {\bf k}_{1} {\bf k}_{2} \ra = \la {\bf x}_{1} | {\bf k}_{1} \ra \la {\bf x}_{2} | {\bf k}_{2} \ra + \la {\bf x}_{2} | {\bf k}_{1} \ra \la {\bf x}_{1} | {\bf k}_{2} \ra.
\eeq
%

%%%%%%%%%%%%%%
\subsubsection{Gaussian quadrature}

As shown above, the single-particle wavefunctions 
[c.f. Eqs.~(\ref{Eq:wf2D}) and~(\ref{Eq:wf3D})] and the associated density functions $n_\beta({\bf x})$, 
$n^{F,B}_\beta({\bf x}_1,{\bf x}_2)$, are governed in the radial variable by a Gaussian decay.
For that reason, it is appropriate to calculate the corresponding integrals using Gauss-Hermite quadrature.
The corresponding $M$ points $x_i$ and $M$ weights $w_i$ allow us to estimate integrals according to
\beq
\int_{-\infty}^{\infty} dx \, e^{-x^{2}} f(x) = \sum_{i=0}^{M-1} w_{i} f(x_{i}).
\eeq
In this work we use the same quadrature points and weights as in our previous work of Refs.~\cite{Casey1, Casey2, Casey3}.

%%%%%%%%%%%%%%%%%%%%%%%%%%%%
\section{Results}
\subsection{Noninteracting virial coefficients at finite angular momentum}

For future reference, and because we have not been able to locate these results elsewhere in the literature, we 
present here the calculation of the noninteracting virial expansion when $\omega_z \neq 0$. We begin with the
well-known result for the partition function of spin-$1/2$ fermions in terms of the single-particle energies $E$:
\beq
\ln \mathcal Z = 2 \sum_{E} \ln \left (1 + z e^{- \beta E } \right),
\eeq
which is valid for arbitrary positive $z$, whereas for (doubly degenerate) bosons
\beq
\ln \mathcal Z = 2 \sum_{E} \ln \left (\frac{1}{1 - z e^{- \beta E }} \right),
\eeq
which is valid for arbitrary $z < \exp(\beta E_0)$, where $E_0$ is the ground-state energy
[$z=\exp(\beta E_0)$ being the well-known limit of Bose-Einstein condensation].
From these expressions, it is easy to see that the virial coefficients $b_n$ for noninteracting 
bosons and fermions differ by a factor of $(-1)^{n+1}$. As is well known, for homogeneous, nonrelativistic
fermions in $d$ dimensions, $b_n = (-1)^{n+1} n^{-(d+2)/2}$. Below, we address the generalization of this 
formula to harmonically trapped systems at finite angular momentum in 2D and 3D.

%%%%%%%%%%%%
\subsubsection{Two spatial dimensions}

In 2D, $E = E_{k m} = \omega_\text{tr} (2k + |m| + 1) + \omega_z m$, where 
$k \geq 0$ and $m$ is summed over all integers.
Thus, we may write the sum by Taylor-expanding the logarithm as
\bea
\ln \mathcal Z &=& 2 \sum_{n = 1}^{\infty} \frac{(-1)^{n+1}}{n} {z^n e^{-n \beta \omega_\text{tr}}} 
\sum_{k=0}^{\infty} e^{- \beta \omega_\text{tr} 2k n } \times \nonumber \\
&&\left[ \sum_{m = 0}^{\infty} e^{- \beta \omega_+ m n } + \sum_{\bar m = 1}^{\infty} e^{- \beta \omega_- \bar m n }\right] .
\eea
where $\omega_\pm = \omega_\text{tr} \pm  \omega_z$.
Carrying out the sums over $k,m,\bar m$, we obtain
\beq
\ln \mathcal Z = Q_1 \sum_{n = 1}^{\infty} b_n z^n,
\eeq
where
\beq
Q_1 b_n = \frac{2\, (-1)^{n+1}}{n} \frac{e^{-n \beta \omega_\text{tr}}} { (1 - e^{- n\beta \omega_+ }) (1 - e^{-n\beta \omega_-})}.
\eeq

Finally, to determine $b_n$ we use $Q_1$ as derived above in Eqs.~(\ref{Eq:Q1_2D}) and~(\ref{Eq:Q1_3D}), such that
\bea
b_n &=& \frac{ (-1)^{n+1}}{n} e^{-\beta \omega_\text{tr} (n-1)}  \frac{ (1 \!-\! e^{- \beta \omega_+}) (1 \!-\! e^{-\beta \omega_-})}
{(1 \!-\! e^{- n\beta \omega_+ }) (1 \!-\! e^{-n \beta \omega_-})}.
\eea
Note that the $b_n$ are always finite, in particular in the  `deconfinement limit' referred to in the introduction where $\omega_- \to 0$, 
\beq
\label{Eq:bnDL2D}
b_n \to b_n^\text{DL2D} \equiv \frac{ (-1)^{n+1}}{n^2} e^{-\beta \omega_\text{tr} (n-1)}  \frac{ (1 - e^{- 2\beta \omega_\text{tr}})}{ (1 - e^{- 2n\beta \omega_\text{tr} })}.
\eeq
On the other hand, $Q_1$ diverges in that limit, because the energy spectrum then becomes independent from $m$.
Simply put, in that limit the centrifugal motion due to rotation 
is strong enough to overcome the trapping potential and the system escapes to infinity. In terms of $\ln \mathcal Z$, the divergence 
may be regarded as a phase transition at $\omega_z = \omega_\text{tr}$. Below we further interpret this limit, considering the 2D and 3D
cases simultaneously. 

We can now derive a virial expansion for the angular momentum and the $z$ component of the moment of inertia:
\beq
\langle \hat L_z \rangle = \frac{\partial \ln \mathcal Z}{\partial (\beta \omega_z)} = Q_1 \sum_{n=1}^{\infty} L_n z^n,
\eeq
where
\beq
\label{Eq:ellnbn2D}
L_n = \frac{1}{Q_1} \frac{\partial \left( Q_1 b_n \right)}{\partial (\beta \omega_z)}  = n b_n \frac{e^{-n \beta \omega_-} - e^{-n \beta \omega_+}} { (1 - e^{-n\beta \omega_+ }) (1 - e^{-n\beta \omega_-})},
\eeq
and
\beq
I_z = \frac{\partial^2 \ln \mathcal Z}{\partial (\beta \omega_z)^2} = Q_1 \sum_{n=1}^{\infty} I_n z^n,
\eeq
where
\bea
\label{Eq:InLn2D}
I_n &=& \frac{1}{Q_1}\frac{\partial (Q_1 L_n)}{\partial (\beta \omega_z)} \\
&=&\!\!-n L_n\!\!
\left[ \frac{e^{-n \beta \omega_+} \!+\! e^{-n \beta \omega_-}} { e^{-n \beta \omega_+} \!-\! e^{-n \beta \omega_-}}
 \!+\! \frac{2 ( e^{-n \beta \omega_+} \!-\! e^{-n \beta \omega_-} ) } { (1 \!-\! e^{-n\beta \omega_+ }) (1 \!-\! e^{-n\beta \omega_-})} \right] 
 \nonumber.
\eea
%
%%
%\bea
%I_n &=& -\frac{1}{Q_1}\frac{\partial (Q_1 \ell_n)}{\partial (\beta \omega_z)} \nonumber \\
%&=&n b_n \times\nonumber \\
%&& \!\!\!\!\!\!\!\!\!\!\!\!\! \left[ \frac{e^{-n \beta \omega_+} + e^{-n \beta \omega_-}} { (1 - e^{-n\beta \omega_+ }) (1 - e^{-n\beta \omega_-})}
% + \frac{2 ( e^{-n \beta \omega_+} - e^{-n \beta \omega_-} )^2 } { (1 - e^{-n\beta \omega_+ })^2 (1 - e^{-n\beta \omega_-})^2} \right].
%\eea
%%

Note that, correctly, $L_n \to 0$ at $\omega_+ = \omega_-$, which corresponds to $\omega_z = 0$, i.e. no rotation.
On the other hand, as may be expected from our previous discussion $L_n \to \infty$ as $\omega_- \to 0$, as in that 
limit the induced rotation overpowers the external potential that holds the system together.
Furthermore, at $\omega_z = 0$, a finite moment of inertia remains:
\bea
I_n \to 2 n (-1)^{n+1} e^{-(2n - 1) \beta \omega_\text{tr}}
\frac{(1 - e^{- \beta \omega_\text{tr}})^2} {(1 - e^{-n \beta \omega_\text{tr}})^4},
\eea
which characterizes the static response to small rotation frequencies within the virial expansion, as a function
of $\beta \omega_\text{tr}$.

%%%%%%%%%%%%
\subsubsection{Three spatial dimensions}

In 3D, $E = E_{k l m} = \omega_\text{tr} (2k + l + 3/2)+ \omega_z m$,
where $k \geq 0$, $l \geq 0 $, and $-l \leq m \leq l$.
Therefore, analyzing the problem as in the 2D case, we obtain
\beq
Q_1 b_n = \frac{2\, (-1)^{n+1}}{n}\frac{e^{-\frac{3}{2}n\beta \omega_\text{tr}} }{(1 \!-\! e^{-n\beta \omega_\text{tr}}) (1 \!-\! e^{-n\beta \omega_{+}}) (1 \!-\! e^{-n\beta \omega_{-}})},
\eeq
and
\bea
b_n &=& \frac{(-1)^{n+1}}{n}e^{-\frac{3}{2}\beta \omega_\text{tr}(n-1)} \times \nonumber \\ 
&& \frac{(1 - e^{-\beta \omega_\text{tr}})(1 - e^{-\beta \omega_+})(1 - e^{- \beta \omega_{-} }) }
{(1 - e^{-n \beta \omega_{\text{tr}}})(1 - e^{- n \beta \omega_+})(1 - e^{- n \beta \omega_{-} })}  .
\eea

As in the 2D case, the $b_n$ are always finite and, in particular in the deconfinement limit $\omega_- \to 0$,
\bea
\label{Eq:bnDL3D}
b_n \to b_n^\text{DL3D} &\equiv& \frac{ (-1)^{n+1}}{n^2} e^{-\frac{3}{2}\beta \omega_\text{tr} (n-1)}  \nonumber  \\
&&\times
\frac{ (1 - e^{- \beta \omega_\text{tr}})}{ (1 - e^{- n\beta \omega_\text{tr} })}
\frac{ (1 - e^{- 2\beta \omega_\text{tr}})}{ (1 - e^{- 2n\beta \omega_\text{tr} })},
\eea
whereas $Q_1$ diverges in that limit.
In this case, the problem can be traced back to the infinite sequence of states for which $\ell = -m$. 
We can also obtain expressions for the virial expansion of the angular momentum and the moment of inertia.
%%
%\beq
%\label{Eq:ellnbn3D}
%L_n = n b_n \frac{e^{-n \beta \omega_-} - e^{-n \beta \omega_+}} { (1 - e^{-n\beta \omega_+ }) (1 - e^{-n\beta \omega_-})},
%\eeq
%%
%
%\bea
%I_n &=& \nonumber
%-n L_n
%\left[ \frac{e^{-n \beta \omega_+} + e^{-n \beta \omega_-}} { e^{-n \beta \omega_+} - e^{-n \beta \omega_-}}
% + \frac{2 ( e^{-n \beta \omega_+} - e^{-n \beta \omega_-} ) } { (1 - e^{-n\beta \omega_+ }) (1 - e^{-n\beta \omega_-})} \right] 
%\eea
%%
Because the dependence of $Q_1 b_n$ on $\omega_+$ and $\omega_-$ is the same in 2D and 3D, 
the relationship between $L_n$ and $b_n$ is identical in 2D and 3D, i.e. Eq.~(\ref{Eq:ellnbn2D}) is valid
in 3D, as long as the $b_n$ corresponding to 3D is used in the right-hand side. Similarly, Eq.~(\ref{Eq:InLn2D}) for $I_n$ carries
over to 3D, as long as the $L_n$ corresponding to 3D is used in the right-hand side.
As expected, and as in the 2D case, $L_n \to 0$ at $\omega_z =0$, whereas 
\beq
I_n \to 2 n (-1)^{n+1} e^{-\frac{1}{2} \beta \omega_\text{tr}(5n - 3)}
\frac{(1 - e^{- \beta \omega_\text{tr}})^3} {(1 - e^{-n \beta \omega_\text{tr}})^5}.
\eeq

\begin{figure}[t]
\includegraphics[width=\columnwidth]{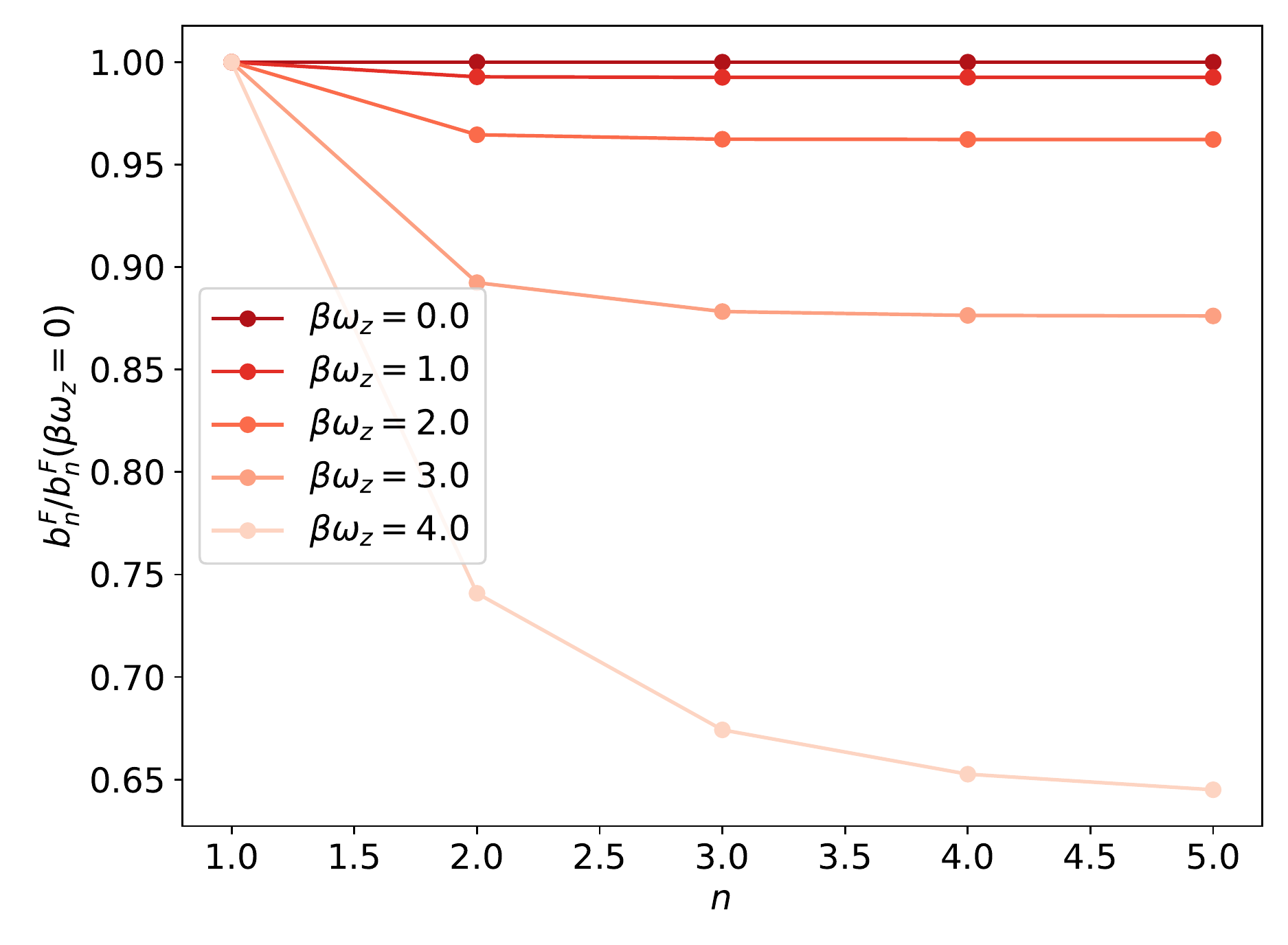}
\caption{\label{Fig:FreeRotatingbn3D} 
Noninteracting $b_n$ normalized by their non-rotating, noninteracting values $b_{n}(\beta \omega_z = 0)$, as functions of $n$ for
a few values of $\beta \omega_z$ and fixed $\beta \omega_\text{tr} = 5$. The ratio $b_n / b_{n}(\beta \omega_z = 0)$ is the same
for bosons and fermions, and is the same in 2D and 3D.
}
\end{figure}

The impact of rotation, i.e. a finite $\beta \omega_z$ on a noninteracting system is displayed in 
Fig.~\ref{Fig:FreeRotatingbn3D}, where we show the ratio of the rotating to non-rotating virial coefficients.
This ratio is the same for bosons and fermions in the noninteracting case and it drastically increases as
$\omega_z$ approaches $\omega_\text{tr}$. At large $n$, this ratio becomes
\beq
\frac{b_n}{b_{n}(\beta \omega_z = 0)} \to 
\frac{(1 - e^{-\beta \omega_+})(1 - e^{- \beta \omega_{-} }) }
{(1 - e^{-\beta \omega_\text{tr}})^2}.
\eeq

\begin{figure}[t]
\includegraphics[width=\columnwidth]{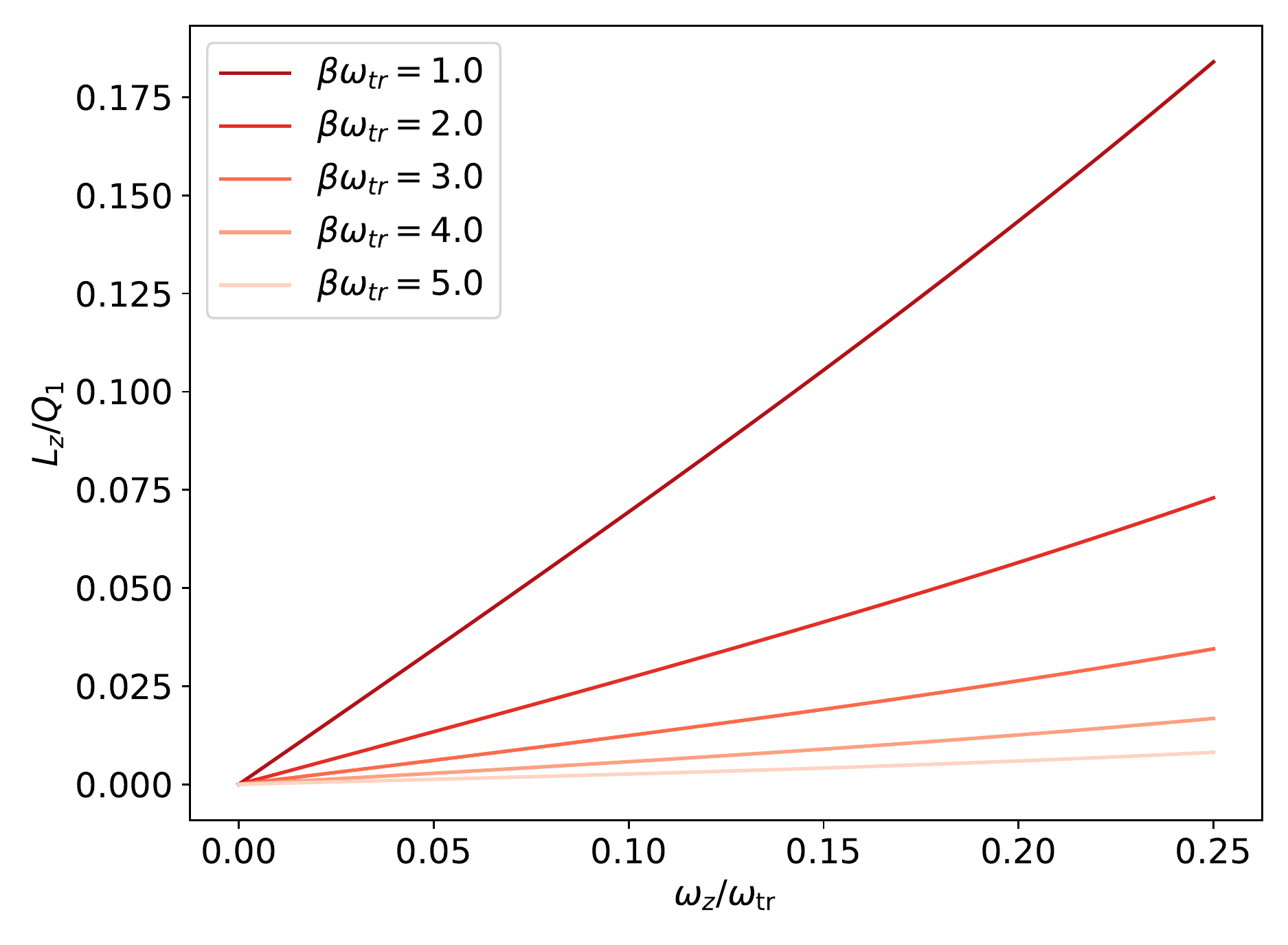}
\caption{\label{Fig:FreeRotatingLz} Noninteracting $L_{z}/Q_{1}$ in for bosons in 3D, as a function of $\omega_{z}/\omega_\text{tr}$ for
a few different temperatures $\beta \omega_\text{tr}$, at third order in the virial expansion.
}
\end{figure}

\begin{figure}[t]
\includegraphics[width=\columnwidth]{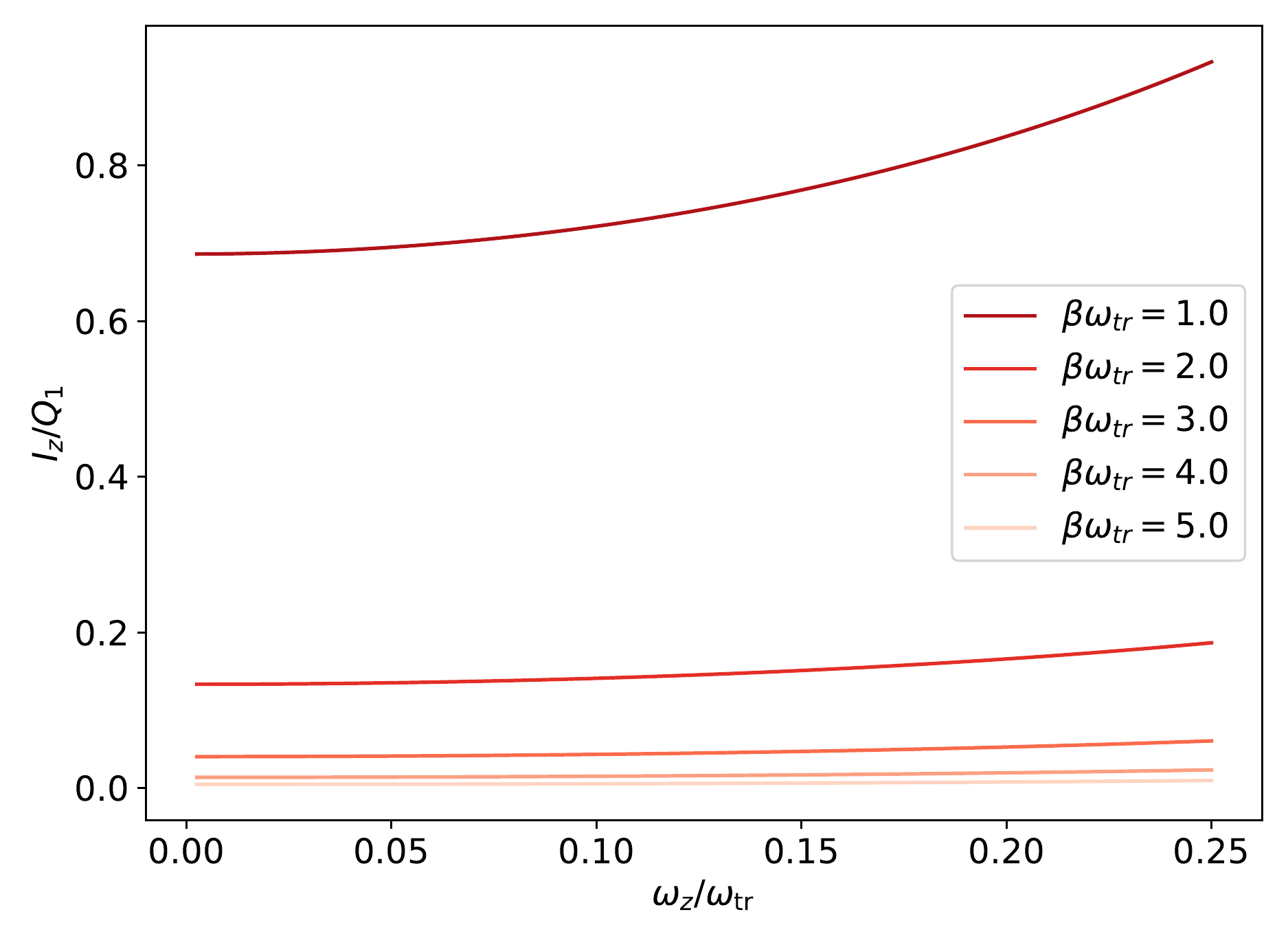}
\caption{\label{Fig:FreeRotatingIz} Noninteracting $I_{z}/Q_{1}$ in for bosons in 3D, as a function of $\omega_{z}/\omega_\text{tr}$ for
a few different temperatures $\beta \omega_\text{tr}$, at third order in the virial expansion.
}
\end{figure}

Naturally, the total angular momentum will increase with $\omega_z$. For a noninteracting system the result is shown in Fig.~\ref{Fig:FreeRotatingLz}
as a function of $\omega_z/\omega_\text{tr}$, at several temperatures $\beta \omega_\text{tr}$. At small $\omega_z$, we find
the linear response regime from which we can extract the moment of inertia $I_z$, as shown in Fig.~\ref{Fig:FreeRotatingIz}.
At the lowest temperatures (highest values of $\beta \omega_\text{tr}$), the response of the system to rotation is highly suppressed,
as seen in both Fig.~\ref{Fig:FreeRotatingLz} and Fig.~\ref{Fig:FreeRotatingIz}. On the other hand, at high temperatures 
(low $\beta \omega_\text{tr}$), where response is higher, we find a mild non-linear regime in which $I_z$ varies
as a function of $\omega_z/\omega_\text{tr}$.

%
%%%%%%%%%%%%%%%%%
\subsubsection{The virial expansion in the deconfinement limit}

Using the limiting expressions for the trapped, rotating $b_n$ in 2D and 3D, namely Eqs.~(\ref{Eq:bnDL2D}) and~(\ref{Eq:bnDL3D}),
respectively, we may analyze the behavior of the system in that limit. To that end, we analyze those equations isolating their 
asymptotic form, which dominates the behavior of the virial expansion series:
\beq
b_n^\text{DL2D} \simeq 2 \frac{ (-1)^{n+1}}{n^2} e^{-\beta \omega_\text{tr} n} \sinh(\beta \omega_\text{tr}),
\eeq
\beq
b_n^\text{DL3D} \simeq 4 \frac{ (-1)^{n+1}}{n^2} e^{-\frac{3}{2}\beta \omega_\text{tr} n}  
\sinh(\beta \omega_\text{tr}/2)\sinh(\beta \omega_\text{tr}),
\eeq
We thus see that the thermodynamics of the deconfined limit is governed in 2D by
\beq
\frac{\ln \mathcal Z}{Q_1} \simeq -2 \sinh(\beta \omega_\text{tr})\; \text{Li}_2 (-e^{-\beta \omega_\text{tr}} z),
\eeq
where $\text{Li}_n(x)$ is the polylogarithm function of order $n$. Similarly, in 3D we obtain
\beq
\frac{\ln \mathcal Z}{Q_1} \simeq -4\sinh(\beta \omega_\text{tr}/2)\sinh(\beta \omega_\text{tr})\; \text{Li}_2 (-e^{-\frac{3}{2}\beta \omega_\text{tr}} z).
\eeq
Notably, and prefactors aside, both the 2D and 3D cases are completely captured by the {\it same} polylogarithm function. 
More specifically, $\text{Li}_2(x)$ is the same function that characterizes the 2D {\it homogeneous} quantum gas (both fermions and bosons).
We therefore see explicitly how, in the deconfined limit, the maximized angular momentum flattens the (3D) system and effectively 
turns it into a homogeneous 2D gas, with a shifted chemical potential. While above we have written the results for fermions, analogous expressions are valid for bosons.

%%%%%%%%%%%%%%%%%%%%%%%%%%%%%%%%%%%
\subsection{Interaction effects on the virial expansion}

In this section we use our results for $\Delta b_2$ and $\Delta b_3$ to calculate the 
angular momentum equation of state, as well as the static response encoded in the moment of inertia.
Denoting the noninteracting grand canonical partition function by $\mathcal Z_0$, we have
\beq
\ln \left(\mathcal Z / \mathcal Z_0\right) = Q_1 \sum_{n=2}^{\infty} \Delta b_n z^n,
\eeq
such that the interaction effect on the angular momentum virial coefficient $L_n$ is
\beq
\Delta L_n = \frac{1}{Q_1} \frac{\partial \left(Q_1 \Delta b_n \right)}{\partial (\beta \omega_z)} = 
\frac{\partial \left(\Delta b_n \right)}{\partial (\beta \omega_z)}
+
\Delta b_n \frac{\partial \left(\ln Q_1 \right)}{\partial (\beta \omega_z)},
\eeq
and its counterpart for the moment of inertia is
\beq
\Delta I_n = \frac{1}{Q_1} \frac{\partial \left(Q_1 \Delta L_n \right)}{\partial (\beta \omega_z)} = 
\frac{\partial \left(\Delta L_n \right)}{\partial (\beta \omega_z)}
+
\Delta L_n \frac{\partial \left(\ln Q_1 \right)}{\partial (\beta \omega_z)},
\eeq
where, using the previous equation for $\Delta L_n$,
\beq
\frac{\partial \left(\Delta L_n \right)}{\partial (\beta \omega_z)} =
\frac{\partial^2 \left(\Delta b_n \right)}{\partial (\beta \omega_z)^2}
+
\frac{\partial \left(\Delta b_n \right)}{\partial (\beta \omega_z)}\frac{\partial \left(\ln Q_1 \right)}{\partial (\beta \omega_z)}
+
\Delta b_n \frac{\partial^2 \left(\ln Q_1 \right)}{\partial (\beta \omega_z)^2}.
\eeq

\begin{figure}[t]
\includegraphics[width=\columnwidth]{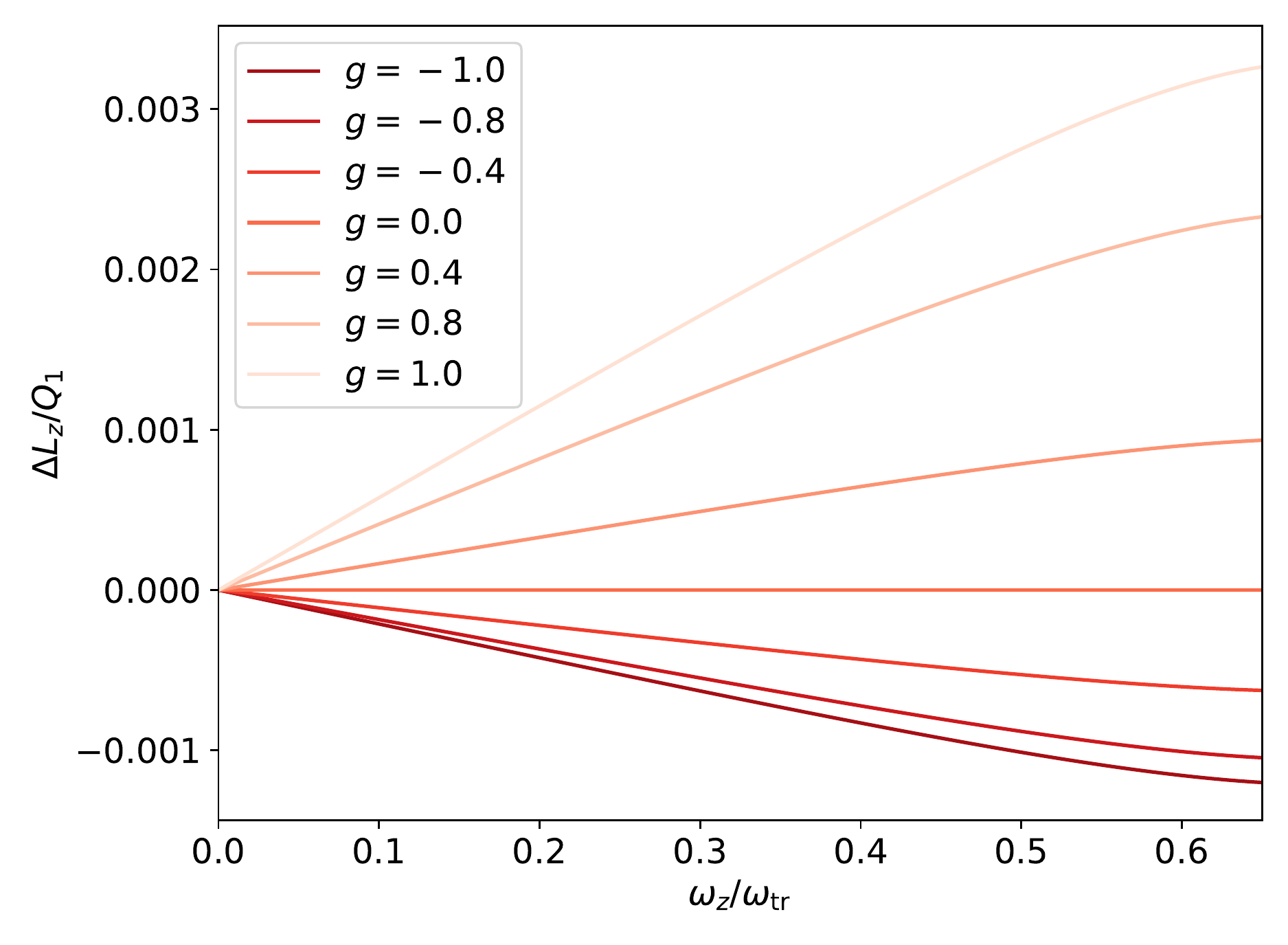}
\caption{\label{Fig:IntRotatingLz3D} Interaction-induced change in the angular momentum of a 3D Fermi gas with attractive
and repulsive contact interactions, as a function of the rotation frequency $\omega_z$ in units of the trapping frequency $\omega_\text{tr}$, at $z = \exp(-2.0)$.}
\end{figure}

\begin{figure}[t]
\includegraphics[width=\columnwidth]{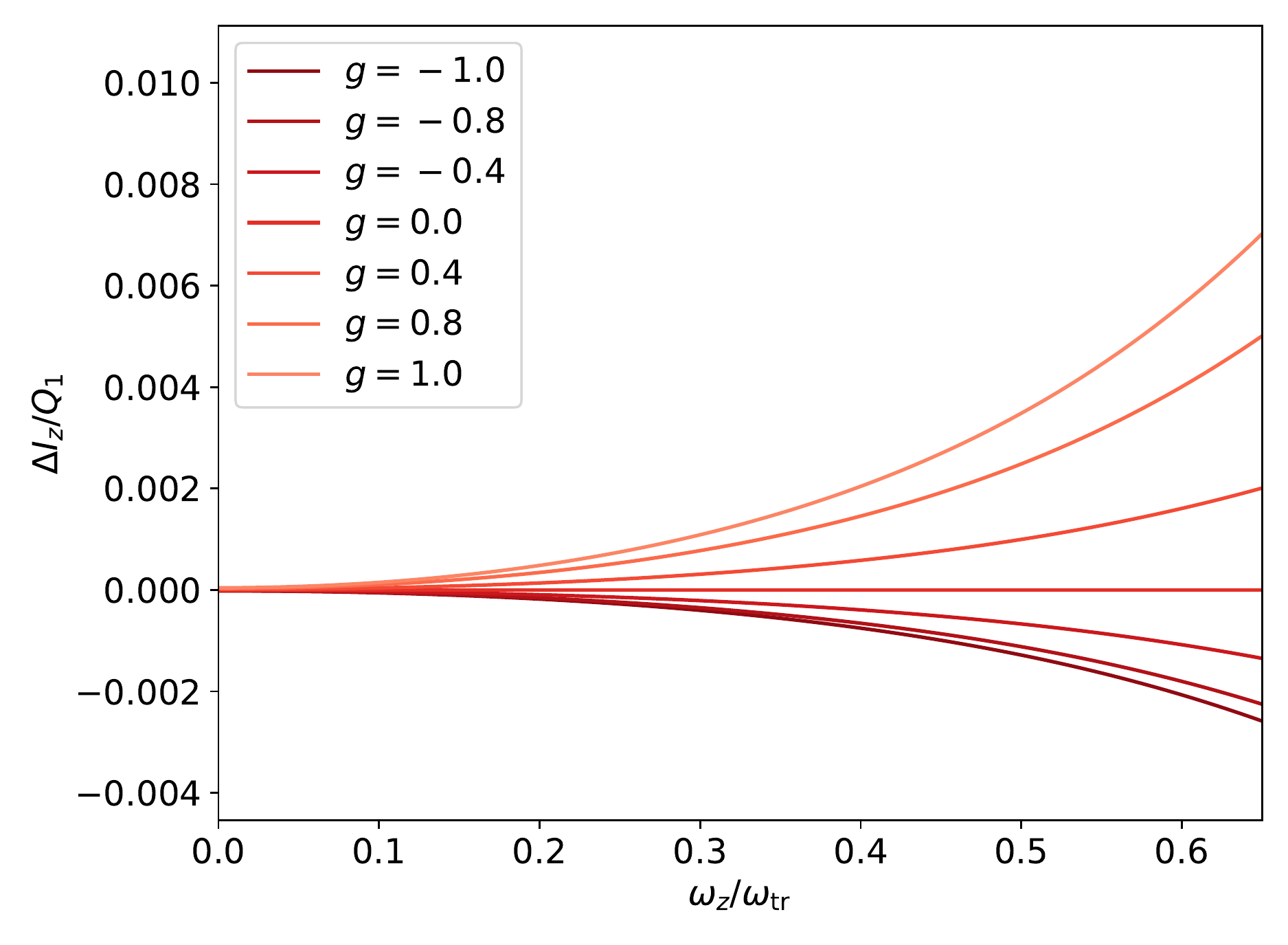}
\caption{\label{Fig:IntRotatingIz3D}  Interaction-induced change in the moment of inertia of a 3D Fermi gas with attractive
and repulsive contact interactions, as a function of the rotation frequency $\omega_z$ in units of the trapping frequency $\omega_\text{tr}$,
at $z = \exp(-2.0)$.}
\end{figure}

Using the above formulas, along with the expressions obtained above for $\Delta b_2$ and $\Delta b_3$ in the
coarse temporal lattice approximation, we readily obtain expressions for the interaction-induced change in the
second- and third-order virial coefficients for the angular momentum and moment of inertia, namely $\Delta L_2$, 
$\Delta L_3$, $\Delta I_2$, and $\Delta I_3$. Based on those, we can rebuild $\langle \hat L_z\rangle/Q_1$ and $I_z/Q_1$
and explore their change due to interactions in the virial region, which we show for fermions in Figs.~\ref{Fig:IntRotatingLz3D} 
and~\ref{Fig:IntRotatingIz3D}. In both figures we find that interactions change the response
to rotation: both the angular momentum and the moment of inertia are modified by correlations,
and the effect increases with $\omega_z$. In particular, attractive
interactions tend to make the system more compact (i.e. they reduce the size of the cloud) thus reducing the moment
of inertia and the total angular momentum, for a given rotational frequency. The corresponding opposite behavior is
found for repulsive interactions. 
%
%While the above simple picture holds for fermions, we observe a qualitatively different behavior for bosons,
%as shown in Figs.~\ref{Fig:IntRotatingLz3DBosons} and~\ref{Fig:IntRotatingIz3DBosons}:
%as repulsive interactions spread out the density distribution (thus increasing $I_z$), they also reduce particle number 
%(thus decreasing $I_z$). For fermions, the former effect dominates, whereas for bosons the latter has the largest impact
%at low rotational frequencies.

%
%

%%
%\begin{figure}[t]
%\includegraphics[width=\columnwidth]{Delta_Lz_F_v_wz_3D}
%\caption{\label{Fig:IntRotatingLz3DBosons} Interaction-induced change in the angular momentum of a 3D Fermi gas with attractive
%and repulsive contact interactions, as a function of the rotation frequency $\omega_z$ in units of the trapping frequency $\omega_\text{tr}$, at $z = \exp(-2.0)$.}
%\end{figure}
%%
%%
%\begin{figure}[t]
%\includegraphics[width=\columnwidth]{Delta_Iz_F_v_wz_3D}
%\caption{\label{Fig:IntRotatingIz3DBosons}  Interaction-induced change in the moment of inertia of a 3D Fermi gas with attractive
%and repulsive contact interactions, as a function of the rotation frequency $\omega_z$ in units of the trapping frequency $\omega_\text{tr}$,
%at $z = \exp(-2.0)$.}
%\end{figure}
%%

%%%%%%%%%%%%%%%%%%%%%%%%%%%%
\section{Summary and Conclusions}

In this work, we have characterized the thermodynamics of rotating Bose and Fermi gases in 2D and 3D using the virial expansion.
To that end, we calculated the effect of rotation on the virial coefficients $b_n$ corresponding to the pressure and density equations 
of state, as well as on the virial coefficients for the angular momentum $L_n$ and moment of inertia $I_n$. We carried out
calculations for interacting as well as noninteracting systems.

In the absence of interactions, we obtained analytic formulas for $b_n$, $L_n$, and $I_n$ in 2D and 3D, which were absent
from the literature to the best of our knowledge. We noted that, while the $b_n$ remain finite when $\omega_z$ approaches 
$\omega_\text{tr}$, the $L_n$, and $I_n$ coefficients diverge, as does the single-particle partition function $Q_1$. 
The origin of the divergence is traced back to the fact that the system becomes unstable at $\omega_z = \omega_\text{tr}$; 
in that deconfinement limit, the high angular velocity enables particles to escape the trapping potential. By exploring
the asymptotic behavior of $b_n$ in that limit, we found that (up to overall factors) it corresponds to that of a homogeneous 
2D gas with a chemical potential shifted by the zero-point energy of the trapping potential.

To address the interacting cases, we implemented a coarse temporal lattice approximation, which allowed us to bypass solving 
the rotating $n$-body problem to calculate the $n$-th order virial coefficient, which we accessed at second and third orders. 
Based on those results, we obtained qualitative estimates for the angular momentum as well as the moment of inertia, as 
functions of the angular velocity $ 0< \omega_z < \omega_\text{tr}$ and temperature $\beta \omega_\text{tr}$. Notably, we find that both the interacting and noninteracting
cases display linear response to rotation at low $\omega_z$, as expected, but we 
are also able to distinguish a non-linear regime in which $I_z$ varies with $\omega_z$;
this is most evident at high temperatures and above $\omega_z/\omega_\text{tr} \simeq 0.1$.

Our work represents a step toward characterizing the properties of rotating matter in high-temperature regimes.
Future studies using increased computational power should be able to explore higher-order corrections
to the coarse lattice approximation presented here. 

This material is based upon work supported by the National Science Foundation under Grant No. PHY1452635 (Computational Physics Program). C.E.B. acknowledges support from the United States Department of Energy through the Computational Science Graduate Fellowship (DOE CSGF) under grant number DE-FG02-97ER25308.

%%%%%%%%%%%%%%%%%%%%%%%%%%%%
\begin{appendix}

%%%%%%%%%%%%%%%%%%%%%%%
\section{Single-particle basis in 2D}

For completeness, in this appendix we show the solution of the Schr\"odinger equation for 
a harmonically trapped particle coupled to the $z$ component of angular momentum in 2D.
The purpose of presenting this information is to establish our notation and to provide a reference point for future work.

We begin with the Schr\"{o}dinger equation in polar coordinates:
\bea
\left(- \frac{\partial^{2}}{\partial r^{2}} - \frac{1}{r} \frac{\partial}{\partial r}- \frac{1}{r^{2}} \frac{\partial^{2}}{\partial \phi^{2}}+ m^{2} \omega_\text{tr}^{2}r^{2} - 2m E\right) \Psi(r,\phi) &=& 0\nonumber
\eea
We then change variables such that $\rho =m \sqrt{\omega_\text{tr}} r$, and $m, \hbar = 1$, which yields
\bea
r &\to& \frac{1}{\sqrt{\omega_\text{tr}}} \rho, \nonumber \\
\frac{\partial}{\partial r} &\to& \sqrt{\omega_\text{tr}}\frac{\partial}{\partial \rho}, \nonumber \\
\frac{\partial^{2}}{\partial r^{2}} &\to& \omega_\text{tr}\frac{\partial^{2}}{\partial \rho^{2}}. \nonumber
\eea
With those replacements, we write $\Psi(\rho, \phi)$ as a product of functions of two individual variables, $\Psi(\rho, \phi) = R(\rho)\Phi(\phi)$, such that
\beq
\left[- \left(\rho^{2}\frac{\partial^{2}}{\partial \rho^{2}} + \rho \frac{\partial}{\partial \rho} + \frac{\partial^{2}}{\partial \phi^{2}}\right)+\rho^{4} - 2\rho^{2}\frac{E}{\omega_\text{tr}}\right]R(\rho)\Phi(\phi)=  0,\nonumber
\eeq
This decouples our partial differential equation into two ordinary equations, each of which must be equal to a constant 
$\widetilde{m}^{2}$:
\bea
- \frac{1}{\Phi(\phi)}\frac{\partial^{2}}{\partial \phi^{2}}\Phi(\phi) =  \widetilde{m}^{2},\nonumber\\
- \frac{\rho^{2}}{R(\rho)}\frac{\partial^{2}R(\rho)}{\partial \rho^{2}} -\frac{ \rho}{R(\rho)} \frac{\partial R(\rho)}{\partial \rho}+\rho^{4} - 2\rho^{2}\frac{E}{\omega_\text{tr}}= -\widetilde{m}^{2}.\nonumber 
\eea
We can solve the equation for $\Phi(\phi)$ straightforwardly: $\Phi(\phi) \propto e^{ i \widetilde{m} \phi}$, with the constraint that $\widetilde{m}$ must be an integer to ensure the solution is not multivalued.

The equation for $\rho$, setting $\widetilde{E} = {E}/{\omega_\text{tr}}$, is then
\bea
- \rho^{2}\frac{\partial^{2}R(\rho)}{\partial \rho^{2}} - \rho \frac{\partial R(\rho)}{\partial \rho}
+ \left(\widetilde{m}^{2} + \rho^{4} - 2\rho^{2}\widetilde{E}  \right) R(\rho) = 0.
\eea

At long distances ($\rho \rightarrow \infty$) we have a harmonic oscillator equation
\beq
-\frac{\partial^{2}R(\rho)}{\partial \rho^{2}} + \rho^{2}R(\rho) =2\widetilde{E}R(\rho),
\eeq
which indicates that at long distances the solution behaves as a Gaussian. 

At short distances ($\rho \ll 1$), on the other hand, our equation reduces to
\beq
-\rho^{2} \frac{\partial^{2}R(\rho)}{\partial \rho^{2}} - \rho \frac{\partial R(\rho)}{\partial \rho} + \widetilde{m}^{2} R(\rho) = 0.
\eeq
We can approach this by proposing proposing $R(\rho) = R_{0} \rho^{c}$, which leads to an equation for the power $c$ in terms of our constant $\widetilde{m}$:
\beq
-c^{2} = \widetilde{m}^{2},\ c = \pm \widetilde{m}.
\eeq
The case $\widetilde{m}=0$ yields two solutions: a constant $R(\rho) = R_{0}$ and $R(\rho) = \ln \rho$. We can discard the second one since it diverges at the origin, which our wave function should not do. For the same reason we discard the case $\widetilde{m} < 0$. Therefore, the short-distance behavior is $R(\rho) \propto \rho^{|\widetilde{m}|}$. 

Based on the above analysis, we propose for the full solution the form: 
\beq
R(\rho) = e^{ -\rho^{2} /2} \rho^{|\widetilde{m}|} F(\rho),
\eeq
where $F(\rho)$ is a function to be determined. This captures the behavior of $R(\rho)$ in our limiting cases. With that form, the radial equation becomes
\beq
\label{Eq:Frho}
\rho^{2} \frac{\partial^{2} F(\rho)}{\partial \rho^{2}} +  \frac{\partial F(\rho)}{\partial \rho}(b_{\widetilde{m}}\rho - 2 \rho^{3}) - 2 a_{\widetilde{m}} \rho^{2} F(\rho) = 0,
\eeq
where $a_{\widetilde{m}} \equiv 1 - \widetilde{E} + |\widetilde{m}|$ and $b_{\widetilde{m}} \equiv 2|\widetilde{m}| + 1$.
We propose a power series form
\beq
F(\rho) = \sum_{k = 0}^{\infty} \rho^{k} c_{k}
\eeq
and obtain algebraic equations for $c_k$ from Eq.~(\ref{Eq:Frho}). Analyzing the lowest powers we obtain the following conditions: From the lowest two powers of $\rho$,
we find that $c_0$ is not fixed but that $c_1 = 0$. The remaining coefficients are related by the recursion
\beq
c_{k+2} = \frac{2(k + a_{\widetilde{m}})}{(k + 2)(k + 1 + b_{\widetilde{m}})} c_k
\eeq
Thus, if both $c_0$ and $c_1$ vanish, then the solution vanishes identically. On the other hand, setting $c_0 = 1$,
only the odd coefficients vanish and we obtain the remaining coefficients recursively. The overall normalization can be
set after the fact since the equation is linear. The series terminates if $k = a_{\widetilde{m}}$ for some $k = 2n \geq 0$ (recall only the even $k$ survive), which yields the quantization condition:
\bea
\frac{E}{\omega_\text{tr}} &=& 2n+|\widetilde{m}|+1.
\eea

\end{appendix}

%%%%%%%%%%%%%%%%%%%%%%%%%%%%

%%%%%%%%%%%%%%%%%%%%%%%%%%%%
\end{document}